\documentclass{article}
\usepackage{graphicx}
\newcommand{\bsigma}{\mbox{\boldmath $\sigma$}}
\newcommand{\btau}{\mbox{\boldmath $\tau$}}
\newcommand{\half}{\frac{1}{2}}
\newcommand{\br}{{\bf r}}

\newcommand{\what}[1]{\widehat #1}
\newcommand{\pu}{p_1}
\newcommand{\pd}{p_2}
\newcommand{\hu}{h_1}
\newcommand{\hd}{h_2}
\begin{document}
\begin{titlepage}
\thispagestyle{empty}
\vspace*{2cm}

\begin{center}

{\Large \bf Correlations and charge distributions of medium heavy nuclei}
\\
\vspace{1.5cm}
{\large Marta Anguiano and Giampaolo Co'} \\
\vspace {.5cm}
{Dipartimento di Fisica, Universit\`a di Lecce  and 
I.N.F.N. sez. di Lecce, \\ I-73100 Lecce, Italy}
\end{center}
\vskip 1.cm 
\begin{abstract}
The effects of long- and short-range correlations on the charge
distributions of some medium and heavy nuclei are investigated. The
long-range correlations are treated within the Random Phase
Approximation framework and the short-range correlations with 
a model inspired to the Correlation Basis Function theory. 
The two type of correlations produce effects of the same order of
magnitude. A comparison with the empirical charge distribution
difference between $^{206}$Pb and $^{205}$Tl shows the need of
including both correlations to obtain a good description of the data. 
\end{abstract} 
\end{titlepage}
\newpage
\section{Introduction}

The high precision of modern electron scattering experiments imposes 
severe constraints to nuclear models and theories.
The study of the charge density distributions of medium and heavy
nuclei by means of elastic electron scattering has allowed for a
detail test of the capacities of the independent particle model (IPM)
to describe the nuclear ground state \cite{cav82}. 

In the IPM the motion of each nucleon inside the nucleus
is not affected by the presence of the other nucleons. 
The corrections beyond this picture take the
generic name of correlations. The sources of correlations can be
generically classified in two categories.
The so called short-range correlations (SRC) are generated by 
the hard core repulsion of the nucleon-nucleon
interaction which prohibits two nucleons to come too close to each
other.  
The long-range correlations (LRC) are produced by collective
excitation modes of the nucleus originated by the part of the nuclear
hamiltonian neglected by the IPM.

This separation in long- and short-range correlation is an artifact of
our approach to nuclear structure usually starting from an IPM basis. 
Exact solutions of the many-body Schr\"odinger equation,
\cite{pud97,wir00} take automatically
into account both kinds of correlations independently from 
their classification.
When approximate or effective theories are used, the role of one or
the other kind of correlation can be emphasized.
The results recently obtained with Fermi Hypernetted Chain (FHNC)
techniques in doubly magic nuclei shows the lack of some correlation  
effects \cite{fab00,mok00,fab01}, probably LRC.
The aim of this article is to estimate the size of the effects
of both kind of correlations on
the charge density distributions of some medium and heavy nuclei.

We estimate the LRC within a Random Phase Approximation (RPA)
approach, and the SRC effects by using a first order
expansion model based on the Correlated Basis Function (CBF)
theory.  We analyze both the quantitative and
qualitative differences between the effects of the two types
of correlations.
The inclusion of both of them is necessary to describe the charge
distribution difference between $^{206}$Pb and $^{205}$Tl. 
The main message of our work is that, in medium heavy nuclei, the two
kind of correlations produce equally important effects beyond the IPM.

The plan of the work is the following one. We briefly present in sects.
\ref{sect:lrc} and \ref{sect:src} the theoretical background of our
treatment of the correlations. In sect. \ref{sect:app} we first make a
detailed description of the inputs of our calculations, then we present
our results on the charge distributions of $^{16}$O, $^{40}$Ca and 
$^{208}$Pb. At the end of the section we show our results regarding
the charge distribution differences between the above mentioned nuclei
and their isotone partners with one proton less. 
Finally we calculate the
charge density difference between the open shell nuclei $^{206}$Pb and
$^{205}$Tl and we compare it with the empirical one. In
sect. \ref{sect:con} we summarize our work and draw our conclusions.  

\section{Long Range Correlations}
\label{sect:lrc}
The RPA many-body states are defined as:
\begin{eqnarray}
&~& | \Psi_N (RPA) > = Q^\dagger_N | \Psi_0 (RPA) > \\
\label{rpa0def}
&~& Q_N | \Psi_0 (RPA) > = 0
\end{eqnarray}
and the operator $Q_N$ satisfies the equations of motions
\begin{equation}
 < \Psi_0 (RPA) |
 \left[ \delta Q_N, \left[H, Q^\dagger_N  \right]  \right]| \Psi_0 (RPA) > 
  = \omega_N
 < \Psi_0 (RPA) |\left[\delta Q_N  , Q^\dagger_N  \right] | \Psi_0 (RPA) >
\end{equation}
for all the possible variations $\delta Q_N$. 
In the above equations we have indicated with $H$ the nuclear
hamiltonian, with $[,]$ the commutator and with $\omega_N$ the
excitation energy. The RPA theory uses:
\begin{equation}
Q^\dagger_N  = \sum_{ph} X_{ph}(N) a^\dagger_p a_h 
                     - Y_{ph}(N) a^\dagger_h a_p 
\end{equation}
where $a^\dagger$ and $a$ are the creation and destruction operators
respectively.
The $X_{ph}(N)$ and $Y_{ph}(N)$ amplitudes are real numbers and are
obtained by solving the RPA equations. 

We are interested in evaluating the mean value of one-body
operators on the RPA ground state defined by eq. (\ref{rpa0def}).
In a second quantization language this can be written as:
\begin{equation}
< {\cal O} > =  
< \Psi_0 (RPA) | 
\sum_{\mu \mu'} <\mu| {\cal O} | \mu'> a^\dagger_\mu a_{\mu'}  
| \Psi_0 (RPA) > 
\end{equation}
where we have indicated with $| \mu>$ a single particle (s.p.) wave
function generated by the mean field basis.
The matrix element 
$< \Psi_0 (RPA)|a^\dagger_\mu a_{\mu'}| \Psi_0 (RPA) >$
is evaluated as described in references \cite{row68,len90} and we obtain
the result:
\begin{equation}
< {\cal O} > =  \sum_h <h| {\cal O} |h> 
\left[1 - \half \sum_N \sum_p | Y_{ph}(N)|^2 \right]
+ \sum_p <p| {\cal O} |p> 
\left[\half \sum_N \sum_h | Y_{ph}(N)|^2 \right]  
\label{opgen}
\end{equation}
where we have indicated with $p$ and $h$ the particle and hole states
respectively. 
The operator defining the occupation number of a
s.p. state $\mu$ is:
\begin{equation}
{\cal N}_\mu = \delta_{\alpha,\mu}
\label{occop}
\end{equation}
where $\alpha=p,h$ in the sums of eq. (\ref{opgen}).
The modifications of the occupation numbers produced by RPA
correlations has been studied in reference \cite{len90}. In this reference
also the contribution of higher order excitations has been evaluated
by means of a Second RPA calculation. 

We are interested in the charge operator:
\begin{equation}
Q(\br) = \sum_i \delta(\br - \br_i) \half\left[ 1 + \tau_3(i) \right]
\label{densop}
\end{equation}
with $\tau_3(i) = \pm 1$ if the particle $i$ is a proton or a neutron
respectively. 

All our calculations are done in a j-j coupled spherical basis. This
means that the s.p. wave functions are expressed as:
\begin{equation}
\phi^t_{nljm}( \br) = 
R^t_{nlj}(r) \sum_{\mu,s} <l \mu \half s | j m > 
             Y_{l,\mu} (\what{r}) \chi_s
\label{spwf}
\end{equation}
where we have indicated with $ < | >$ the Clebsch-Gordan coefficients,
with $Y_{l,\mu} (\what{r})$ the spherical harmonics, with $\chi_s$
the spin wave function and with $t$ the third component of the
isospin. 

Using the angular momentum coupled RPA equations and the expression
(\ref{spwf}) for the s.p. states, we obtain the following expression
for the charge density distribution of a doubly-closed shell nucleus:
\begin{eqnarray}
\nonumber
4 \pi \rho_{LRC}(r) &=& 
\sum_{(nlj)h} (2j_h+1) (R_{(nlj)h}(r))^2 
\left[1 - \half \, \frac{1}{2j_h+1} 
\sum_p \sum_{J,N} (2J+1) | Y_{ph}(J,N)|^2 \right] \\
&+& \sum_{(nlj)p} (2j_p+1) (R_{(nlj)p}(r))^2 
\left[\half \, \frac{1}{2j_p+1} 
\sum_h \sum_{J,N} (2J+1) | Y_{ph}(J,N)|^2 \right]
\label{rholrc}
\end{eqnarray}
where $p$ and $h$ indicate particle and hole states respectively  
limiting the range of the sums. In our calculations the sums on $p$
and $h$ are restricted only to proton states. We neglect the
contribution of the neutrons to the charge distribution. 
In the IPM the $Y_{ph}(J,N)$ amplitudes are zero and
eq. (\ref{rholrc}) gives the well known expression of $\rho_{IPM}$.  

In this article we calculate also the charge distributions of nuclei
with one proton less than the doubly magic ones.  This is done within
Landau-Migdal theory \cite{mig67} which allows to calculate the
difference between expectation values of operators in doubly magic
nuclei and nuclei with one nucleon less (or more).  The formalism has
been widely used to calculate magnetic moments of nuclei around
$^{208}$Pb \cite{spe77} and it has been extended in references
\cite{co86,co87} to the calculation of the charge distributions.  The
basic equation to be used to evaluate this difference is:
\begin{eqnarray}
\nonumber
&~& <A| {\cal O}_J | A > - <A-1 ; i | {\cal O}_J | k ;A-1> = 
< i || {\cal O}_J || k > \\
\nonumber
&~& + \sum_N  \sum_{\pu \pd \hu \hd} <i \pu || V || k \hu >
\frac {X_{\pu \hu}(J,N) X^*_{\pd \hd}(J,N)} 
{\epsilon_{\pu} - \epsilon_{\hu} - \omega_N} < \pd || {\cal O}_J || \hd > \\
&~& - \sum_N  \sum_{\pu \pd \hu \hd} <i \pu || V || k \hu >
\frac {Y_{\pu \hu}(J,N) Y^*_{\pd \hd}(J,N)} 
{\epsilon_{\pu} + \epsilon_{\hu} - \omega_N} < \hd || {\cal O}_J || \pd > 
\label{ffs}
\end{eqnarray}
In the above equation $V$ indicates the residual interaction used in
RPA calculations, $\epsilon$ the s.p. energy, $\omega_N$ the RPA
excitation energy of the N-th state and $| i >$ and $| k >$ the s.p.
states characterizing the transition in the A-1 system within the
IPM.  To calculate the difference between ground state charge
distribution we consider $J$=0 and the fact that $| i >$ = $| k >$ is
the last occupied proton s.p. level just below the Fermi surface.
\section{Short Range Correlations}
\label{sect:src}
\begin{figure}
%
%
\includegraphics[bb=-250 50 500 680,angle=0,scale=0.4]
{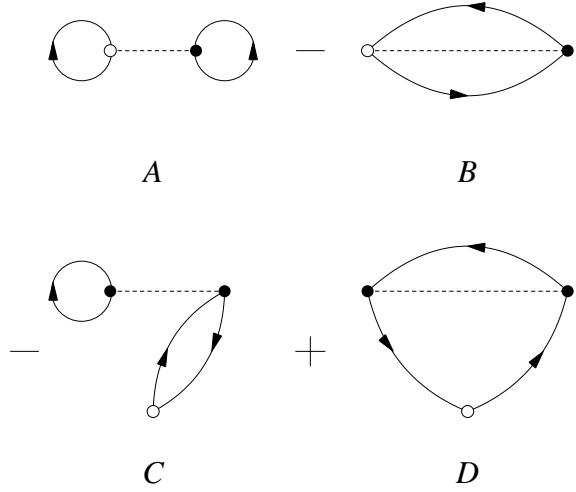}
\vspace {-2.0cm}
\caption{ Cluster diagrams contributing to the one-body
  density. The dashed lines represent the dynamical correlations,
  $h^p$, and the oriented lines the statistical correlations,
  $\rho_0(\br_i, \br_j)$. A black dot associated with a point implies
  integration over its coordinates. 
}
\label{fig:diag}
\end{figure}
We evaluate the 
effects of SRC on the ground state charge distribution by using  
the approach of references \cite{co95,ari97}. 
The starting point is the following ansatz on the ground state:
\begin{equation}
\Psi_0(1,2...A)= G(1,2...A)\Phi_0(1,2...A)
\label{cbf0def}
\end{equation}
where we have indicated with $\Phi_0$ the Slater determinant of
s.p. states. The correlation operator is written as a
symmetrized product of two-body operators:
\begin{equation}
G(1,2...A)={\cal S}\left[\prod_{i<j}F_{ij}\right]
\label{mcor}
\end{equation}
The expression of $F_{ij}$ used in the most sophisticated
calculations, both in nuclear matter and in finite nuclear systems,
$F_{ij}$ has the form:
\begin{equation}
F_{ij}=\sum_{p=1,6}f^p(r_{ij})O^p_{ij}
\label{corrf}
\end{equation}
where
\begin{equation}
O^{p=1,6}_{ij}=
\left[ 1, \bsigma_i \cdot \bsigma_j, S_{ij},\right] \otimes
\left[ 1, \btau_i \cdot \btau_j \right] 
\label{oper}
\end{equation}
In the above equations we have indicated with
$S_{ij}=(3\,{\hat  {\bf r} }_{ij} \cdot \bsigma_i  \,
{\hat{\bf r}}_{ij} \cdot 
\bsigma_j -  \bsigma_i \cdot \bsigma_j)$ the tensor operator. 

The expectation value of a generic  operator ${\cal O}$ is:
\begin{equation}
< {\cal O} > = \frac { <\Psi_0 | {\cal O} |\Psi_0 >} { <\Psi_0 |\Psi_0 >}  
\label{mean}
\end{equation}
In CBF theory the ground state is found through the
minimization of the hamiltonian expectation value. 
The search for this minimum involves
variations on the parameters of the correlation functions 
$f^p$ and of the mean field potential defining the s.p. basis. 

The charge distribution is obtained by inserting in eq.(\ref{mean})
the charge density operator (\ref{densop}). 
We found convenient to rewrite the correlation function (\ref{mcor})
as: 
\begin{equation}
G(1,2...A)=1+{\cal S}  \prod_{i<j}
\left[ \sum_{p=1,6}h^p(r_{ij})O^p_{ij} \right]
\, . 
\label{mcor1}
\end{equation}
where we have defined $h^1=f^1-1$ and $h^{p>1}=f^{p>1}$. 
The expectation value is evaluated with standard cluster expansions
techniques \cite{cla79} which allow us to express it as
infinite sum of linked diagrams. The unlinked terms of the numerators
are canceled by the denominator.

At this point we introduce an approximation. We evaluate only those
diagrams containing a single correlation function $h^p$. 
The diagrams considered are shown in figure \ref{fig:diag}. 
It is evident from Eq. (\ref{mcor1}) that the first term of our
calculation is the IPM density which is already properly normalized.
This implies that in the normalization
procedure the contribution of all the correlation diagrams should
vanish. It is shown in reference \cite{co95} that
the set of diagrams of figure \ref{fig:diag} provides the 
correct normalization of the density.

The validity of our approximation in calculating the charge
distribution has been studied in reference \cite{co95} by comparing
the results of the model with those of a Fermi Hypernetted Chain
calculation. A test of the approximation in the evaluation of the
nuclear matter charge responses has been done in reference
\cite{ama98}.  More recently \cite{fab01} the approximation has been
tested in the evaluation of the one-body densities and the natural
orbits.  The conclusion of these investigations is that the
approximation is reliable. This is probably related to the fact that
the charge operator is a relatively simple one-body operator,
therefore not very sensitive to the fine details of the many-body wave
function.

The calculation of the correlated density is carried on by expanding the
correlation function (\ref{mcor1}) in terms of Legendre
polynomials. This facilitates the calculation in the spherical basis
of the s.p. wave functions (\ref{spwf}). Details are given in
reference \cite{co95}. 

\section{Applications}
\label{sect:app}
\subsection{The input}
\label{subs:input}
We have conducted our study in three different mass regions located
around the doubly magic nuclei $^{16}$O, $^{40}$Ca and $^{208}$Pb.  In
our calculations we used the same set of s.p. wavefunctions for
evaluating both LRC and SRC effects. These wavefunctions have been
generated by diagonalizing a Woods--Saxon potential in a harmonic
oscillator basis. This procedure automatically discretizes the
continuum.  In the RPA calculations we used configuration spaces
considering all the states below the Fermi surface and three major
shells above it. We found good convergence of our calculations when
multipole excitations up to J=10 were used.

%
%
\begin{table}[h]
\begin{center}
\begin{tabular}{l|cccccc}
\hline
 & $f^{in}$ & $f^{ex}$ & $f'^{in}$ & $f'^{ex}$ & $g_0$ & $g'_0$  \\
\hline
 $V_{LM}$ & 0.20 & -2.45 & 1.50 & 1.50 & 0.55 & 0.70 \\
 $V_{JS}$ & 0.21 & -1.80 & 0.65 & 1.65 & 0.70 & 0.70 \\
\hline
\end{tabular}
\end{center}
\caption{Parameters of the Landau-Migdal force, $V_{LM}$, and of the
 zero--range part of the J\"ulich-Stony Brook interaction $V_{JS}$. 
 We used the same normalization constant $C_0$
 for both interaction. For $^{16}$O and $^{208}$Pb calculations 
 we used $C_0$=386 MeV fm$^3$, while for $^{40}$Ca 
 $C_0$=212 MeV fm$^3$. 
 The density dependent parameters have been changed for each nucleus 
 considered. They are $\alpha$=0.50, 0.53, 0.60 fm$^{-1}$ and 
 $R_F$= 3.0, 4.1, 7.22 fm respectively for $^{16}$O, $^{40}$Ca and 
 $^{208}$Pb.
}
\label{tab:force}
\end{table}

The parameters of the potential we have adopted are taken form the
literature \cite{rin78,co87b} and have also been used in the FHNC
calculations of reference \cite{ari96}, where j-j coupling and isospin
dependence of the s.p. wave functions were considered.  We obtained
the charge distribution by folding the pointlike proton distributions
with the nucleon electromagnetic form factors of reference\cite{ber72}.

The RPA calculations used to evaluate the LRC effects require an
additional input: the particle-hole interaction. We have tested the
sensitivity of our results to the choice of this input by using two
different types of interaction. A first one is a Landau-Migdal (LM)
contact interaction of the form:
\begin{equation}
V_{LM} (\br_1,\br_2)= 
C_0 \left[ f_0(r_1) + f'_0(r_1) \btau_1 \cdot \btau_2 
+  g_0(\rho) \bsigma_1 \cdot \bsigma_2 
+  g'_0(\rho)\bsigma_1 \cdot \bsigma_2 \btau_1 \cdot \btau_2 \right]   
\delta^3(\br_1 -\br_2)
\end{equation}

The experience of RPA calculations in the Pb region \cite{spe77} 
has shown the need of imposing a density dependence to the
scalar and isospin parameters $f_0$ and $f'_0$ parameters. 
As suggested in reference \cite{rin78} for these two channels we use the 
expression: 
\begin{equation}
f(r)= f^{ex}+(f^{in}-f^{ex}) \frac{1}{1+exp[\alpha(r-R_F)]}
\end{equation}
where $\alpha$ and $R_F$ are two free parameters. The factor $C_0$
represents the inverse of the density of states at the Fermi surface
and we consider it as a free parameter used to set the full strength
of the force.

The second RPA interaction is the finite range J\"ulich--Stony Brook
(JSB) force \cite{spe80}, obtained by adding to the zero range piece
of Landau-Migdal type, the contribution of one-pion plus one-rho meson
exchange potentials.  This is a finite range interaction and in the
discrete excitation spectrum is necessary to obtain a good description
of the magnetic excitations \cite{co90}.

The values of the parameters used in our calculations are given in
table \ref{tab:force}.  The parameters of the delta interaction are
those of reference \cite{rin78} where have been fixed to reproduce the
excitation energy of the low lying 2$^+$ and 3$^-$ states in
$^{208}$Pb as well as the position of the the monopole giant
resonance.  We have adjusted the free parameters of the JSB force to
reproduce the 3$^-$ state of $^{208}$Pb at 2.61 MeV. We obtained for
the $B(E3;0 \rightarrow 3)$ values: $7.536 \times 10^5$ e$^2$fm$^6$
with LM interaction and $7.489 \times 10^5$ e$^2$fm$^6$ with JSB, in
agreement with the experiment~\cite{boh75}.

Even though the parameters of the two interactions have been tuned in
the lead region their performances in reproducing the $^{16}$O spectrum
are quite good.  The low lying 3$^-$ state of $^{16}$O was obtained at
the energy of 6.95 MeV and at 7.35 MeV respectively for the LM and the
JSB interactions.  These values should be compared with the
experimental values of 6.13 MeV. In this case the $B(E3;0 \rightarrow
3)$ values are $0.97 \times 10^2$ e$^2$fm$^6$ for LM and $1.17 \times
10^3$ e$^2$fm$^6$ for JSB to be compared with the experimental value,
$1.5 \times 10^3$ e$^2$fm$^6$~\cite{boh75}. This good
result is obtained because the empirical single particle energies are
rather well reproduced by the spherical Woods-Saxon potential.

The situation is rather different for $^{40}$Ca.  In this case the
splitting between the single particle levels around the Fermi surface
is narrower than the empirical one. For this reason the interaction
fixed in $^{208}$Pb is too attractive. We got an imaginary eigenvalue
for the first 3$^-$ state.  We reduced the value of $C_0$ from $386$
MeV fm$^3$ down to $212$ MeV fm$^3$ and we obtained for the 3$^-$ the
energies of 1.48 MeV and 3.20 MeV for LM and the JSB forces
respectively. The experimental value is 3.73 MeV. The LM force gives a
too low $B(E3;0 \rightarrow 3)$ value, $1.1762 \times 10^4$
e$^2$fm$^6$ (the experimental one is $1.70 \times 10^4$
e$^2$fm$^6$~\cite{boh75}), while the JSB gives $1.22 \times 10^4$
e$^2$fm$^6$.

%
%
\begin{figure}
\includegraphics[bb=-60 50 500 680,angle=0,scale=0.7]
{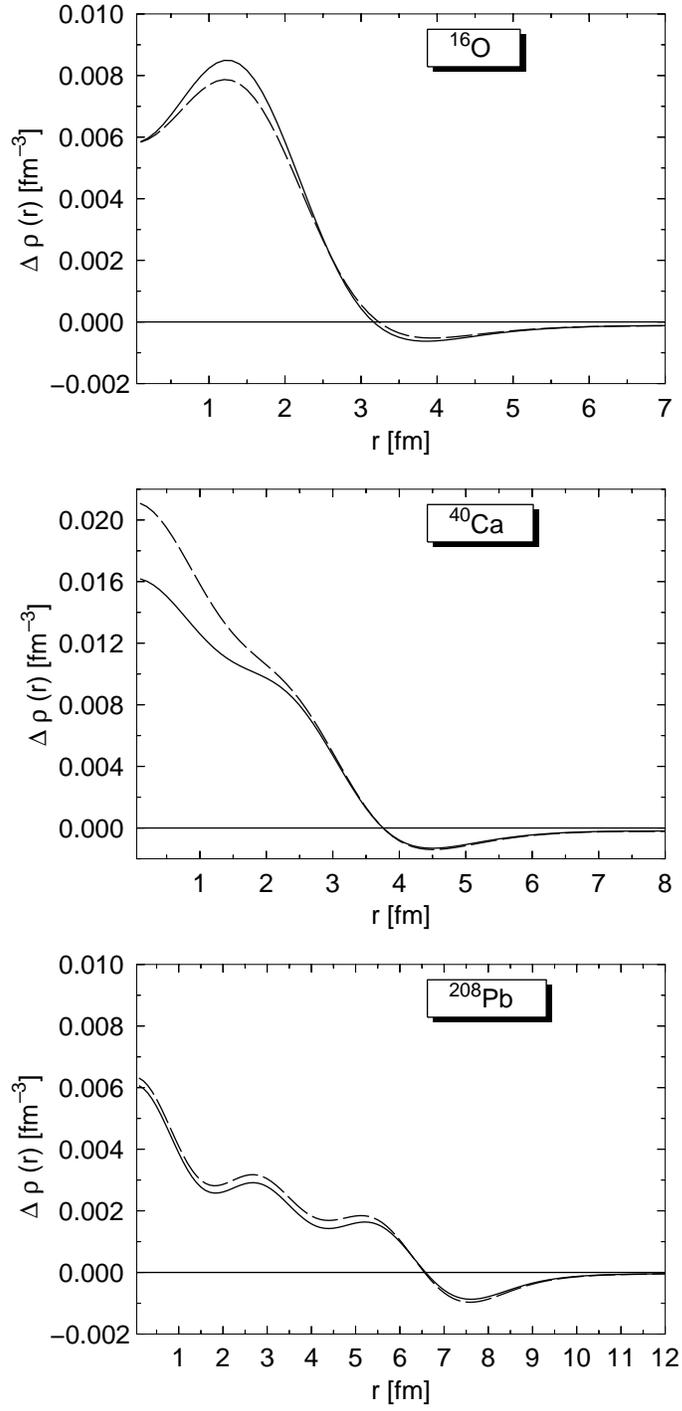}
\vspace{0.5cm} 
\caption{ Differences between IPM charge densities and those obtained 
  considering LRC correlations. The full lines have been
  obtained with the J\"ulich-Stony Brook interaction, the dashed lines
  with the  Landau-Midgal force. 
}
\label{fig:lrcdiff}
\end{figure}
%
%
\begin{figure}
\includegraphics[bb=-60 50 500 680,angle=0,scale=0.7]
{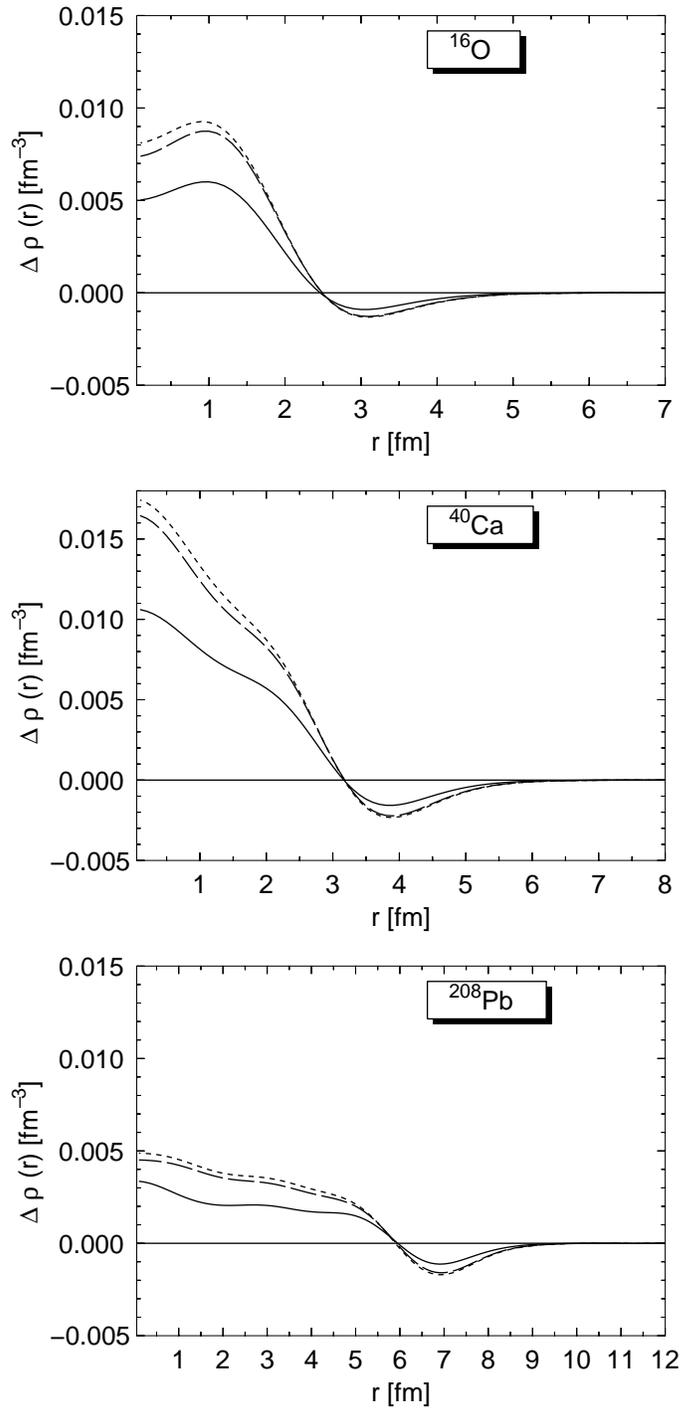}
\vspace{0.5cm}
\caption{
  Differences between IPM densities and those obtained considering SRC
  correlations. The full lines have been obtained with the state
  dependent correlations of reference  \protect\cite{fab00} and the long
  dashed lines by using only the central term of this correlation.
  The gaussian correlation of reference \protect\cite{ari96} produces the
  short dashed lines.  }
\label{fig:srcdiff}
\end{figure}
In figure \ref{fig:lrcdiff} we show the differences between the IPM
charge distributions and those calculated considering the LRC
evaluated with the two interactions. 
We can see that the densities calculated with the Landau--Migdal and  
the J\"ulich-Stony Brook interactions are very similar. 
The differences between the results obtained with the two forces
are  within the accuracy of the data.

In our model, the input required to evaluate the SCR effects are the
correlation functions of eq.(\ref{mcor1}). In variational
calculations these functions, together with the s.p. wave functions, 
are determined by the minimization procedure. In the  
calculations of reference \cite{ari96} where the semi-realistic
nucleon-nucleon S3 interaction \cite{afn68} has been used
simple scalar correlation functions were required. 
In our calculations we have considered the gaussian correlations of
reference \cite{ari96}. Their expression is:
\begin{equation}
f^{(1)}(r)=1-A \exp (- B r^2) 
\end{equation}
and the parameters $A$ and $B$ have been fixed by the minimization
procedure. It has been shown in reference \cite{ari96} that, for a fixed 
nucleon-nucleon interaction, the correlation functions 
are almost the same for every nucleus considered. 
The obtained values of the two parameters were $A$=0.7
and $B$=2.2 fm$^{-2}$. The differences between IPM charge distributions
and those calculated with this correlation function are given by the
short dashed-lines of figure \ref{fig:srcdiff}.

We studied the effects of the state dependent terms of the correlation
function by using the correlations defined by the FHNC calculations of
reference \cite{fab00}. In this reference the realistic V8'
nucleon-nucleon interaction plus the Urbana IX three-nucleon
interaction \cite{pud97} have been used.
The difference between IPM charge distributions and those obtained
with this correlation functions are presented in
figure \ref{fig:srcdiff} by the full lines. We performed calculations
switching off the state dependent part of the correlations and we used
only the scalar part. The results of these calculations are given by
the long dashed lines of figure \ref{fig:srcdiff}. 
When the state dependent part of the correlation is switched off we
essentially reproduce the curves obtained with the gaussian
correlation. The contribution of the state dependent correlations
diminish the difference with the IPM results. 
These results confirm the findings of references \cite{fab01,co95,ari97}. 

In the calculations we shall discuss in the next sub-sections 
we shall refer to the LRC calculations as those done with
the J\"ulich--Stony Brook interaction and to the SRC as those done
with the full state dependent correlation of reference \cite{fab00}.

\subsection{Doubly magic nuclei}
\label{subs:magic}
\begin{figure}
\includegraphics[bb=-60 50 500 680,angle=0,scale=0.7]
{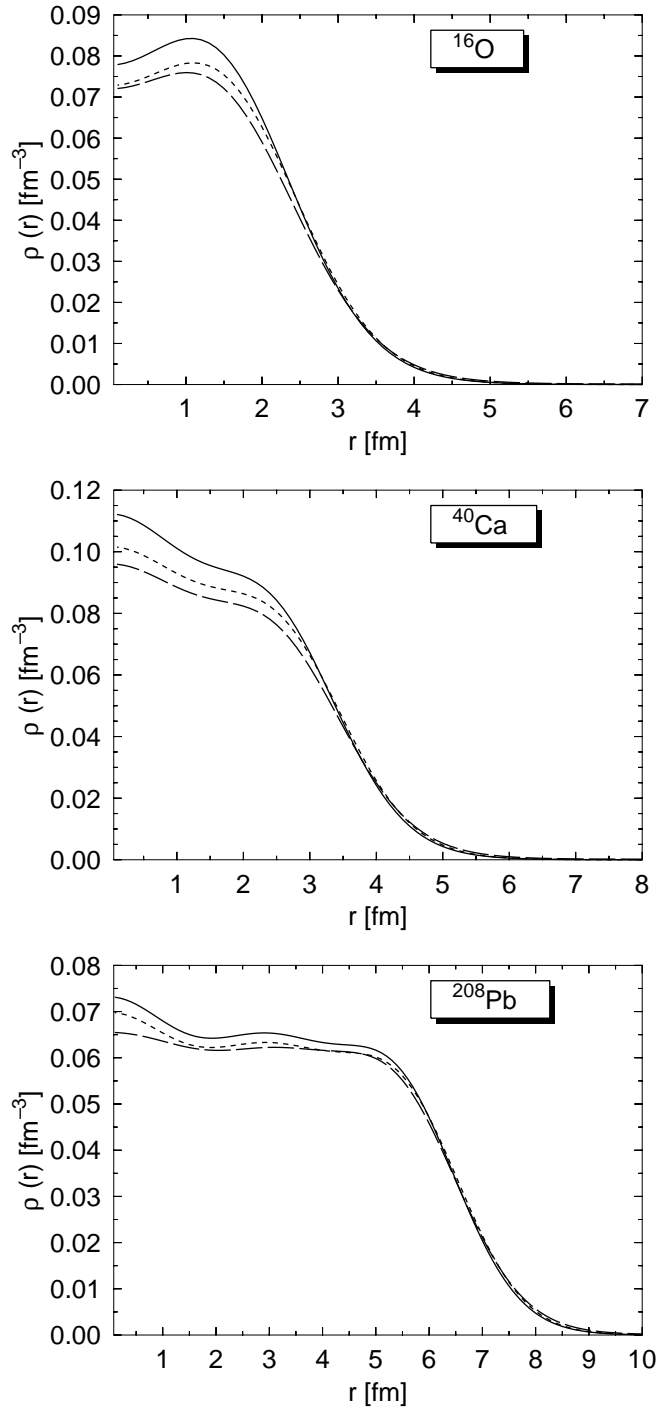}
\vspace{0.5 cm}
\caption{ Charge density distributions for the three closed shell
  nuclei investigated. IPM (full lines), 
  LRC (long dashed lines), SRC (short dashed lines). 
}
\label{fig:dens}
\end{figure}

The IPM charge distributions of $^{16}$O, $^{40}$Ca and $^{208}$Pb
are compared in figure \ref{fig:dens} with those obtained 
after the inclusion of the correlations. 
The effects produced by both kind of correlations are analogous.
The charge distribution is lowered in the center of the nucleus and 
this implies an increase of the rms charge radii as it is shown in
table \ref{tab:rms}. 

%
%
\begin{table}[h]
\begin{center}
\begin{tabular}{l|c|c|c}
\hline  & {$^{16}$O} 
        & {$^{40}$Ca} 
        & {$^{208}$Pb} \\
\hline
 IPM & 2.64 & 3.25  & 5.46 \\
 SRC & 2.70 & 3.31  & 5.52 \\
 LRC & 2.98 & 3.57  & 5.59 \\
\hline
\end{tabular}
\end{center}
\caption{ Root mean square charge radii in fm.}
\label{tab:rms}
\end{table}

\begin{figure}
\includegraphics[bb=-60 50 500 680,angle=0,scale=0.7]
{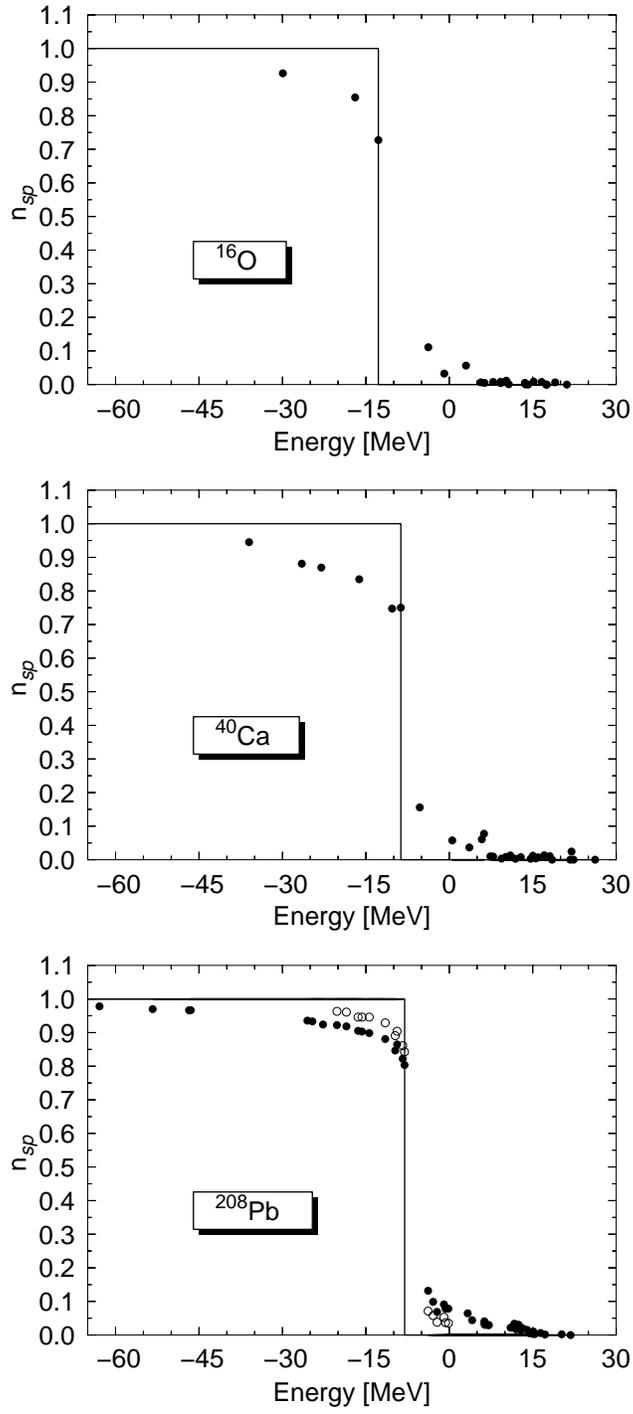}
\vspace{0.5 cm}
\caption{
  Proton occupation numbers as a function of the single particle
  energy. The full lines represent a Fermi gas distribution.  The
  white dots in the lowest panel show the results of reference
  \protect\cite{gog79}. }
\label{fig:octot}
\end{figure}

The correlations change the occupation probability of the
s.p. states with respect to those of the IPM. 
For the LRC we calculated the occupation
probabilities by evaluating the expectation values of the operator
defined in Eq. (\ref{occop}). 
The results are shown in figure \ref{fig:octot} where we
observe that the main changes are around the Fermi surface. 

%
%
%
%
%
%
\begin{figure}
\includegraphics[bb=-60 50 500 680,angle=0,scale=0.7]
{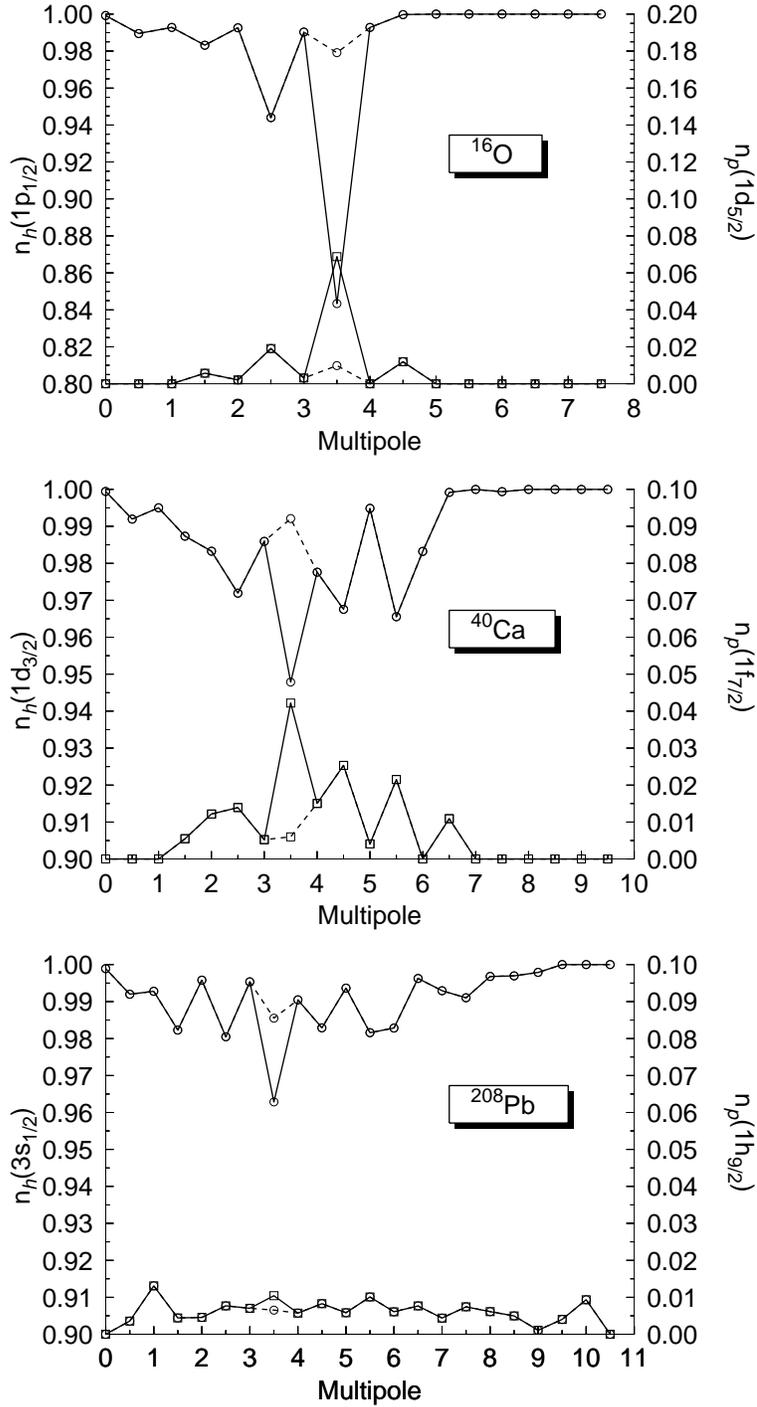}
\vspace{0.5 cm}
\caption{
  Occupation numbers of the valence hole and particle states as a
  function of the excitation multipole. On the abscissa the multipoles
  are odered for a fixed value of $J$ first with positive and then
  negative parity. The left vertical scales and the upper lines refer
  to the hole states and the right scales associated with the lower
  lines to the particle states.  The dashed lines show the results
  obtained by leaving out the lowest collective 3$^-$ states.  }
\label{fig:ocmul}
\end{figure}

Our LRC correlations produce slightly stronger effects than those found
by Gogny \cite{gog79}, whose results are shown by the white dots
in the $^{208}$Pb panel of figure \ref{fig:octot}. 
Our LRC effects are strong also in comparison with those of the RPA
calculations of reference \cite{len90} done in the $^{40}$Ca nucleus. These
differences are due to the different interactions and s.p. bases
used. 

In the LRC calculations the change of the occupation probability is
due to the coupling of the s.p. level with the collective excitation
of the nucleus. This is clearly shown in figure \ref{fig:ocmul} where
the occupation probabilities of the two s.p. levels just below and
above the Fermi surface are shown as a function of the multipole
excitation. It is evident that the occupation probability is strongly
modified in correspondence to the 3$^-$ states. The dashed lines show
the result of a calculation when the lowest collective 3$^-$ state is
not considered.

The SRC correlations produce analogous effects on the occupation
probabilities of the s.p. levels. An evaluation of these modifications
is not straightforward as in the case of LRC since in the CBF
calculations the quantities used are the one- and two-body densities
and not the s.p. wave functions.   
We have calculated the one-body densities for the three nuclei
considered within our first order correlation model by
using the multipole decomposition technique described in 
references \cite{co95,ari97}.
We projected the $l$ multipole term $\rho_l(r_1,r_2)$
of the one-body density matrix on the s.p. basis by calculating
\begin{equation}
n_{SRC}= \int dr_1 \, r^2_1 \int dr_2 \, r^2_2
          R_{nlj}(r_1)  R_{nlj}(r_2) \rho_l(r_1,r_2) 
\end{equation}
The occupation numbers for the $s$ waves obtained with this procedure
are compared in table \ref{tab:occ} with the LRC ones. In our
calculation the LRC are more effective in lowering the occupation
probability of the valence states than the SRC.

%
%
\begin{table}[h]
\begin{center}
\begin{tabular}{l|cc|cc|cc}
\hline  &\multicolumn{2}{c|} {$^{16}$O} 
        &\multicolumn{2}{c|} {$^{40}$Ca} 
        &\multicolumn{2}{c} {$^{208}$Pb} \\
        & LRC  & SRC  & LRC & SRC & LRC & SRC \\
\hline  
 1s1/2  & 0.93 & 0.95 & 0.93 & 0.94 & 0.97 & 0.93 \\
 2s1/2  &      &      & 0.78 & 0.94 & 0.92 & 0.91 \\
 3s1/2  &      &      &      &      & 0.80 & 0.92 \\
\hline
\end{tabular}
\end{center}
\caption{ Occupation probability of the $s$ states.}
\label{tab:occ}
\end{table}

We obtained for the occupation probability of the 3s1/2 state of
$^{208}$Pb the value of 0.74, which is slightly larger than the 0.6
extracted from the (e,e'p) data of ref. \cite{qui86} even though
recent DWBA re-analysis obtained a value of 0.7 \cite{udi93}.  The
nuclear matter studies of SRC and LRC of the occupation probability
predict values around 0.8 \cite{pan84,ben90}. As expected our value is
lower than this since we also include surface vibrations.

\subsection{Semi-magic nuclei}
\label{subs:smagic}
\begin{figure}
\includegraphics[bb=-60 50 500 680,angle=0,scale=0.7]
{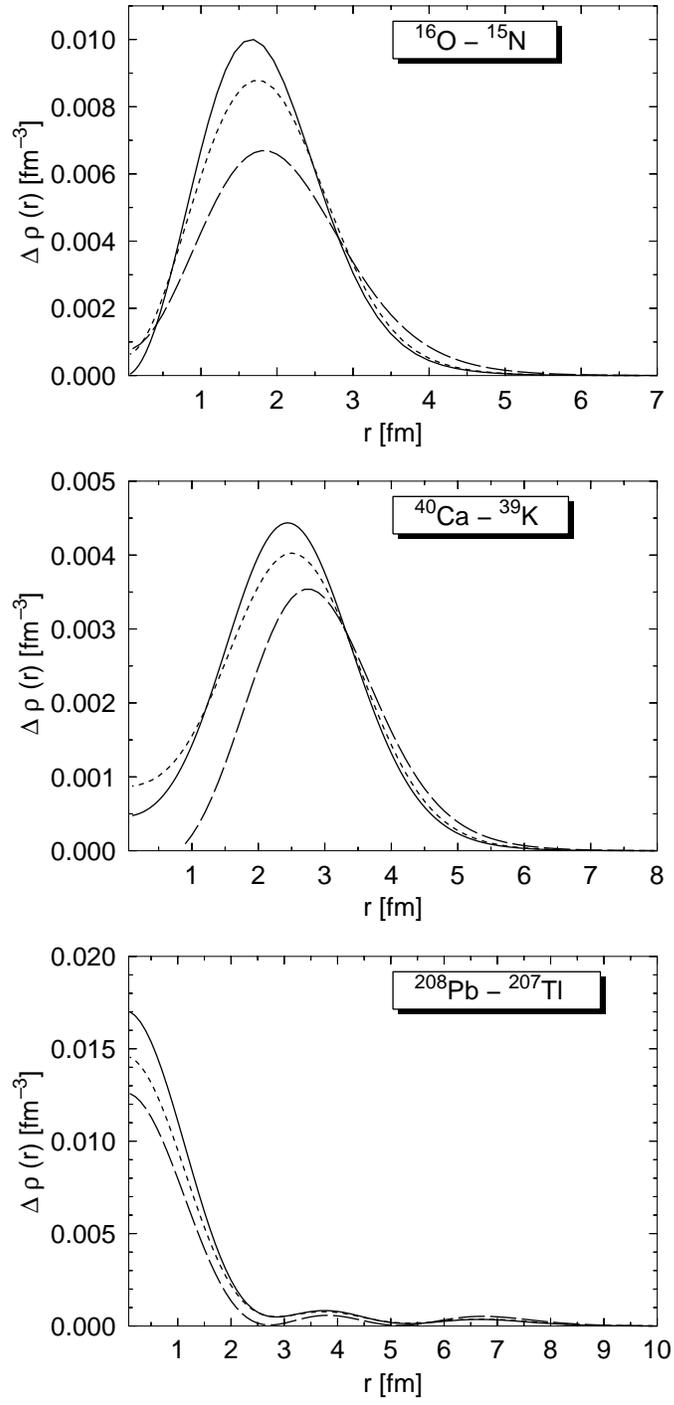}
\vspace {0.5cm}
\caption{Charge distribution differences between isotones.
         IPM (full lines), SRC (short
         dashed lines) and LRC (long dashed lines).  
}
\label{fig:diffA-A1}
\end{figure}

We have calculated the differences between the charge distributions of
isotones differing by one proton only, since we expect this quantity to
be less dependent from the choice of the s.p. wave functions than the
full charge distribution. 
The results of these calculations for the three pairs of isotones
considered are shown in figure \ref{fig:diffA-A1}.
The IPM results correspond to the square of the s.p. wave
function of the least bound proton state.
They are the $1p_{1/2}$, $1d_{3/2}$ and $3s_{1/2}$ s.p. states for the 
$^{16}$O,  $^{40}$Ca and $^{208}$Pb nuclei respectively.
The shape of the curves  shown in the figure reproduce the
known behaviour of these s.p. wave functions. Only the $s$ wave is
peaked in the center of the nucleus, the other ones have their
maxima at larger distances. 

With respect to the IPM results, the correlations diminish the
difference in the central part of the nucleus and tends to shift the
charge in the external regions. 
In the $^{16}$O--$^{15}$N and $^{40}$Ca--$^{39}$K systems this effect 
produces a shift of the maxima.
In the $^{208}$Pb--$^{207}$Tl system the inclusion of the correlations
lowers the difference in the center of the nucleus.
As in the case of the doubly magic nuclei short and long range
correlations produce effects of the same order of magnitude. 

%
%
%
%
\begin{figure}
\includegraphics[bb=-40 50 500 680,angle=0,scale=0.7]
{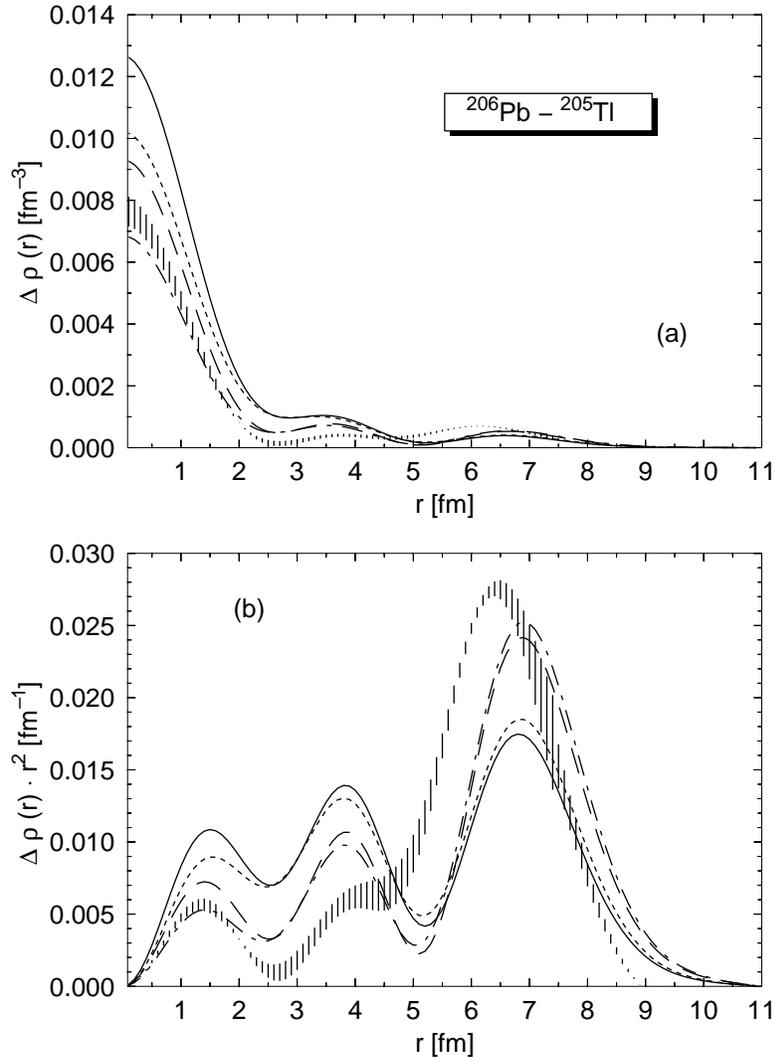}
\vspace{-1.0cm}
\caption{
  Differences between $^{206}$Pb and $^{205}$Tl charge densities. Full
  lines IPM, dotted lines SRC, dashed lines LRC, dashed dotted lines
  full calculation. In the lower panel the charge densities are
  multiplied by r$^2$ to emphasize the large distance behaviour. The
  experimental data are from \protect\cite{cav82,fro83}.  }
\label{fig:diff205}
\end{figure}

The interest in the calculation of these charge differences is
related to the existence of experimental data in the lead region.
In a single experiment \cite{cav82,fro83} the elastic electron
scattering cross sections of $^{206}$Pb and $^{205}$Tl have been
measured. From the analysis of these data the empirical difference
between the charge distributions of these two nuclei has been
extracted. In the center of the nucleus the IPM predicts a
larger charge difference than the one deduced by the experimental
data. 

A direct comparison between the differences
shown in the lowest panel of figure \ref{fig:diffA-A1} and the
experimental one is not correct since the $^{206}$Pb and $^{205}$Tl
nuclei have a rich open shell structure which is not considered in the
$^{208}$Pb and $^{207}$Tl system. 
To describe the two open shell nuclei we adopted the model already
used in  reference \cite{co87}.
The ground state of $^{205}$Tl is described as:
\begin{eqnarray}
\nonumber
| 1/2^+, ^{205} {\rm Tl}> &=& 
      \alpha_1 (|3 s _{1/2}>^{-1} \otimes \, |0^+,^{206}{\rm Pb}> \\ 
\nonumber 
      &+&\alpha_2 (|2 d _{3/2}>^{-1} \otimes \, |2^+,^{206}{\rm Pb}> \\  
      &+&\alpha_3 (|2 d _{5/2}>^{-1} \otimes \, |2^+,^{206}{\rm Pb}> 
\end{eqnarray}
where we considered $\alpha_1$=0.86, $\alpha_2$=-0.47,
$\alpha_3$=0.30, obtained in reference \cite{zam75} using first order
perturbation theory with a $\delta$ interaction.  The $0^+$ and $2^+$
states of the $^{206}$Pb are obtained as linear combination of
two-hole wave functions in the closed neutron core of $^{208}$Pb
\cite{kle76}.  We use the hole-hole amplitudes calculated in reference
\cite{kuo71} within a Tamm--Dankoff Approximation.  The ground state
amplitudes are rather similar to those given in reference \cite{boh75}
which reproduce the empirical values obtained by an analysis of (d,p)
reactions on $^{206}$Pb target.  The IPM charge difference obtained
with these wavefunctions is represented by the full lines of figure
\ref{fig:diff205}.

While the LRC have been consistently treated within 
the open shell structure of the two nuclei \cite{co87}, 
this has not be done for the SRC. In this case we supposed their
effect to be the same as that calculated in closed shell nuclei:
\[
\rho_{\rm SRC}^{206}(r) \simeq \rho_{\rm SRC}^{208}(r) - 
[\rho_{\rm IPM}^{208}(r) - \rho_{\rm IPM}^{206}(r)]   
\] 
and analogously for the 205--207 system. 
With this hypothesis the total charge difference between $^{206}$Pb
and $^{205}$Tl is:
\[
\Delta_{\rm TOT}^{206-205}(r) = 
\Delta_{\rm LRC}^{206-205}(r)+\Delta_{\rm SRC}^{206-205}(r)
-\Delta_{\rm IPM}^{206-205}(r)  
\simeq 
\Delta_{\rm LRC}^{206-205}(r)+\Delta_{\rm SRC}^{208-207}(r)
-\Delta_{\rm IPM}^{208-207}(r)  
\]
where we have indicated with $\Delta$ the difference between the
charge densities of the two nuclei indicated in the upper index.  The
total difference evaluated in this way is compared in figure
\ref{fig:diff205} with the empirical one.  The inclusion of both long
and short range correlations greatly improves the agreement with the
experiment. In the lower panel we show the differences multiplied by
$r^2$ to emphasize the comparison at large values of the nuclear
radius.  Also in this region the inclusion of both correlations
improves the agreement.

\section{Summary and Conclusions}
\label{sect:con}
%
%
\begin{figure}
\includegraphics[bb=-40 50 500 680,angle=0,scale=0.7]
{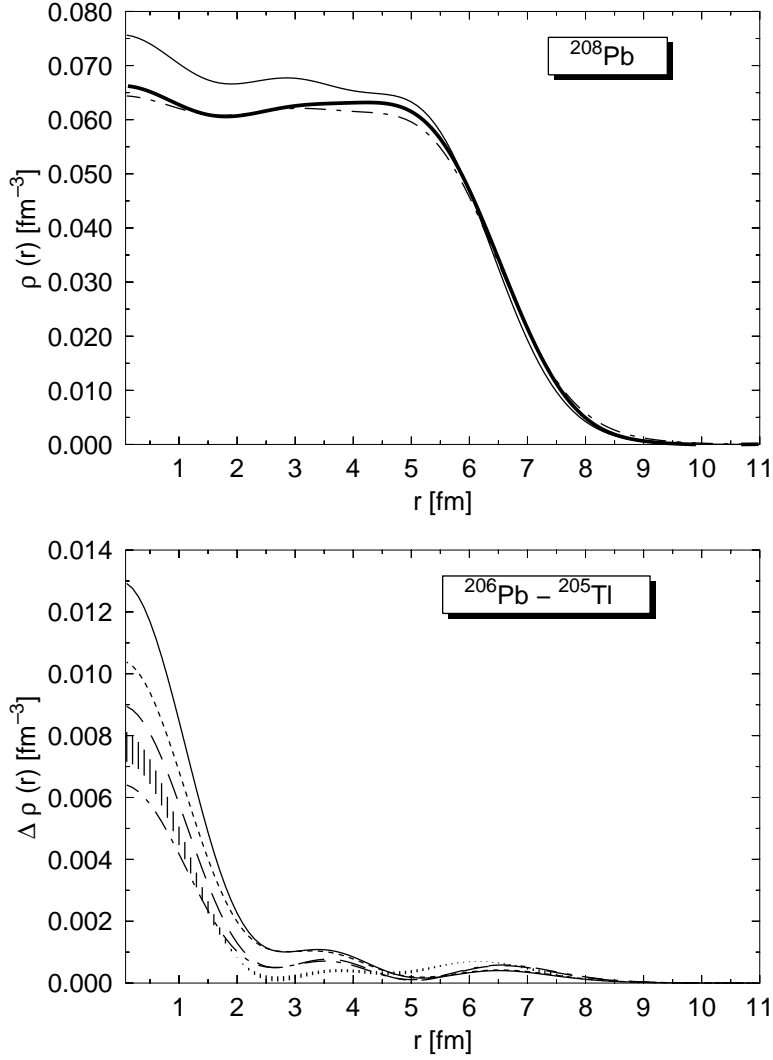}
\vspace {-1.0cm}
\caption{Results obtained with a new set of s.p. states (see text).
  Upper panel: $^{208}$Pb charge distributions compared with the
  empirical one \protect\cite{dej87} represented by the thick line.
  The thin full line show the IPM result, the dashed line has been
  obtained by summing long and short range correlations. Lower panel:
  Differences between $^{206}$Pb and $^{205}$Tl charge densities. The
  meaning of the lines is analogous to that of figure \ref{fig:diff205}.
  }
\label{fig:new208}
\end{figure}
%
%
%
\begin{figure}
\includegraphics[bb=-30 50 500 680,angle=0,scale=0.7]
{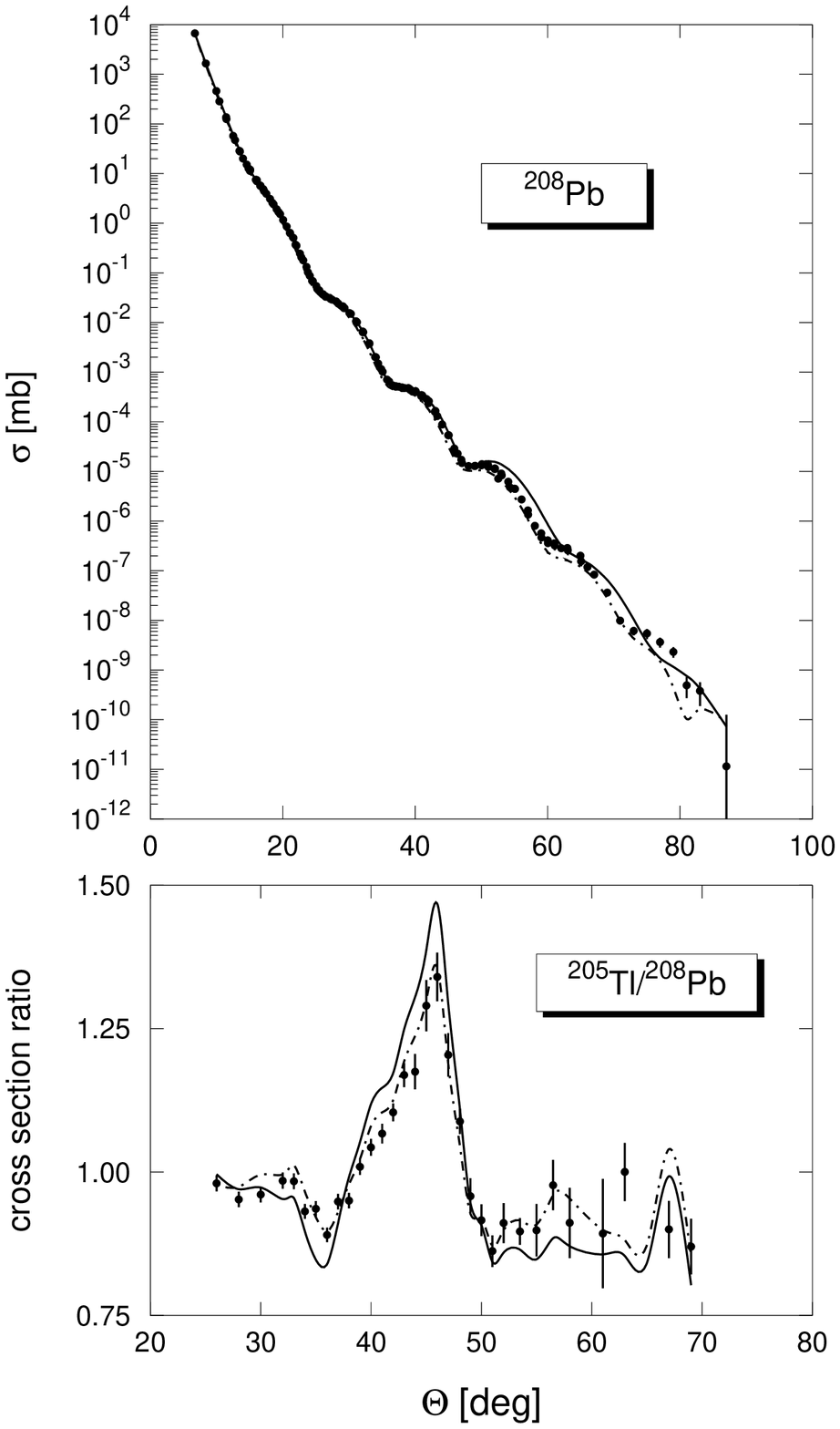}
\vspace {-2.0cm}
\caption{Above: elastic electron scattering
  cross section on $^{208}$Pb (data from \protect\cite{hei69}).
  Below: ratio beteen the elastic cross sections on $^{205}$Tl 
  and $^{206}$Pb targets (data from  \protect\cite{cav82}).
  The full lines represent the IPM results and the dashed-dottoed ones
  the results obtained includibg both LRC and SRC. 
  }
\label{fig:xsect}
\end{figure}

We have calculated the effects of long and short-range correlations on
the charge distribution of some doubly magic nuclei. The SRC have been
treated in the framework of the Correlated Basis Function theory.  We
use a first order model in the correlation function. The correlations
have been taken from FHNC calculations done with realistic
interaction.  The LRC correlations have been treated within a RPA
framework. In this case we used an effective interaction whose
parameters have been fixed to reproduce the $^{208}$Pb low-energy
excitation spectrum.

We have seen that long- and short- range correlations produce similar
effects on the charge distributions. In both cases the IPM
distributions are lowered in the center of the nucleus and, as a
consequence, the rms radii increase. In terms of s.p. occupation
numbers this effect is produced because both kinds of correlations
decrease the occupation of the hole states and increase that of the
particle states. 
We have shown that in all the nuclei considered the link between
s.p. levels and collective low-lying 3$^-$ state
is one of the major sources of modification of the occupation number. 

The charge differences with nuclei having one proton less have been
calculated in order to compare our results with the measured empirical
difference in the Pb region. Also on the charge difference the two
kind of correlations produce similar results. Also in this case the
correlations tend to diminish the difference in the central region. 
After a proper consideration of the open shell structure of the 
$^{206}$Pb and $^{205}$Tl nuclei, the total difference which considers 
both correlations can reproduce the empirical one.

Our calculations do not treat LRC and SRC on the same
ground. A mixing of ingredients taken from fundamental theory, the
correlation functions, and effective theory, the RPA correlations, has 
been used in our calculations. 
In all our work the choice of s.p. wave functions is essential.  
The sets of s.p. states we have used have been taken from the
literature and the parameters of the mean field used to generate them
have been fixed to reproduce at best the s.p. energies of the
valence states around the Fermi level, and the charge distributions.  
For this reason we did not make a comparison between
our charge distributions and the experimental ones. Since the IPM
densities are already reproducing rather well the empirical ones,
the inclusion of correlations worsen the agreement with the
experiment.  

It is sufficient a small change in the mean field parameters to obtain
a set of s.p. states which can reproduce the empirical densities after
the inclusion of the correlations. We have done that for $^{208}$Pb
where we have changed the Woods-Saxon depth $V_0$ from 60.4 MeV to
62.0 MeV and the radius from 7.46 fm to 7.4 fm.  The results obtained
with this new set of s.p. wave functions are shown in figure
\ref{fig:new208}. In the upper panel we compare $^{208}$Pb charge
densities with the empirical one \cite{dej87}.  In the lower panel we
show that even with the new set of s.p. states the charge density
difference between $^{206}$Pb and $^{205}$Tl is well reproduced.

In the upper panel of figure \ref{fig:xsect} we compare with the data
of ref. \cite{hei69} the electron scattering elastic cross sections
calculated within the Distorted Wave Born Approximation 
by using the densities of the previous figure. The
dashed-dotted line obtained by including both SRC and LRC reproduces
better the data than the IPM density (full line).  In the lower panel
we compare our results with the ratio of the elastic cross
sections measured on $^{206}$Pb and $^{205}$Tl targets \cite{cav82}.
Also in this case the line better reproducing the data is the
dashed-dotted one obtained by including all the correlation effects.

We have presented these results to show that we would be able to
describe the empirical densities with small changes of the mean field 
parameters. This was not the goal of our work. We simply wanted to
investigate the relative role played by long and short range
correlations, and we found that they should be
simultaneously considered in order to obtain a successful description
of the empirical densities. This is our main message. 

The facts triggering our investigation are related to
the difficulties found by recent CBF variational calculations
\cite{fab00} in reproducing the experimental binding energies and
charge distributions of some doubly closed shell nuclei. 
These microscopic calculations, depending only from the realistic 
nucleon-nucleon interaction, provide a good description of the
SRC. Our work adds another piece of evidence that the
adequate treatment of the LRC is the missing ingredient. 
Hopefully, perturbative corrections, like those developed in
in references \cite{fan84,ben89} for nuclear matter, 
could provide the required description of the LRC.

\vskip 2.0 cm
{\bf Acknowlegments}\\
We thank P.F. Bortignon and A.M. Lallena for useful discussions.
This work has been partially supported by MURST through the  
{\sl Progetto di Ricerca di Interesse Nazionale:
Fisica teorica del nucleo atomico e dei sistemi a molticorpi}.
%
%
\newpage

\begin{thebibliography}{99}
%
\bibitem{cav82} 
 Cavedon J M {\sl et al} 1982 
 {\sl Phys. Rev. Lett.} {\bf 49} 978 
\bibitem{pud97} 
 Pudliner B S, Pandharipande V R,
 Carlson J, Pieper S C and Wiringa R B 1997 
 {\sl Phys. Rev. } C {\bf  56} 2261 
\bibitem{wir00} 
  Wiringa R B, Pieper S C, Carlson J and
  Pandharipande V R 2000 
  {\sl Phys. Rev. }C {\bf 62} 0144001
\bibitem{fab00} Fabrocini A, Arias de Saavedra F and Co' G 2000 
  {\sl Phys. Rev. }C {\bf 61} 044302 
\bibitem{mok00} 
  Mokhtar S R, Co' G and Lallena A M 2000 
  {\sl Phys. Rev. }C {\bf 62} 067304
\bibitem{fab01} 
  Fabrocini A and Co' G 2001 
  {\sl Phys. Rev. }C {\bf 63} 044319 
\bibitem{row68} 
  Rowe D J 1968 
  {\sl Phys. Rev.} {\bf 175} 1283
\bibitem{len90} 
  Lenske H. and Wambach J 1990 
  {\sl Phys. Lett.} B {\bf 249} 377 
\bibitem{mig67} 
  Migdal A B 1967 {\sl Theory of finite
    Fermi systems and applications to atomic nuclei} 
   (New York: John Wiley) 
\bibitem{spe77} 
   Speth J, Werner E and Wild W 1977 
   {\sl Phys. Rep.} {\bf 33} 127 
\bibitem{co86} 
   Co' G and Speth J 1986 
   {\sl Phys. Rev. Lett.} {\bf 57} 547 
\bibitem{co87} 
   Co' G and Speth J 1987 
   {\sl Zeit. Phys. }A {\bf 326} 361 
\bibitem{co95} 
   Co' G {\sl Nuov. Cim. 1995} A {\bf 108} 623 
\bibitem{ari97} 
   Arias de Saavedra F, Co' G and Renis M M 1997 
   {\sl Phys. Rev. }C {\bf 55} 673
\bibitem{cla79}
  Clark J W 1979 {\sl  Prog. Part. and Nucl. Phys.}  {\bf 3} 89\\
  Pandharipande V R and Wiringa R B 1979
  {\sl Rev. Mod. Phys.}  {\bf 51} 821 \\
  S. Rosati 1982 {\em From nuclei to particles}, {\sl Proc. Int.
  School E. Fermi, course LXXIX} ed. A. Molinari (Amsterdam: North
  Holland) pp 73-112 
\bibitem{ama98} 
  Amaro J E, Lallena A M, Co' G and Fabrocini A 1998 
  {\sl Phys. Rev.} C {\bf 57} 3473 
\bibitem{rin78}
  Rinker G A and Speth J 1978 {\sl Nucl. Phys.} A {\bf 306} 360
\bibitem{co87b} 
  Co' G, Lallena A M and Donnelly T W 1987 
  {\sl Nucl.  Phys.} A {\bf 469} 684 
\bibitem{ari96} 
  Arias de Saavedra F, Co' G, Fabrocini A and Fantoni S 1996 
  {\sl Nucl. Phys.} A {\bf 605} 359
\bibitem{ber72} 
  Bertozzi W, Friar J, Heisenberg J and Negele J W 1972
  {\sl Phys. Lett.} B {\bf 41} 408 
\bibitem{spe80} 
  Speth J, Klemt V, Wambach J and Brown G E 1980 
  {\sl Nucl. Phys.} A {\bf 343} 382
\bibitem{co90} 
  Co' G and Lallena A M 1990 
  {\sl Nucl. Phys. } A {\bf  510} 139 
\bibitem{boh75}
   A. Bohr and B. Mottelson 1975
   {\sl Nuclear Structure}, Vol. II (London: Benjamin) 
   pp 561 and 642
\bibitem{afn68} 
  Afnan I R and Tang Y C 1968 
  {\sl Phys. Rev.} {\bf 175} 1337 
\bibitem{gog79} 
  Gogny D 1979 {\sl Nuclear
  Physics with Electromagnetic Interactions}, ed H. Arenh\"ovel and
  D. Drechsel, Lecture Notes in Physics, Vol. 108 (Berlin: Springer) p
  88 
\bibitem{qui86} 
  Quint E N M {\sl et al} 1986
  {\sl Phys. Rev. Lett.} {\bf 57} 186 \\
  Quint E N M {\sl et al} 1987 
  {\sl Phys. Rev. Lett.} {\bf 58} 1088
\bibitem{udi93} 
  Ud\'{\i}as J M, Sarriguren P, Moya de Guerra E,
  Garrido E and Caballero J A 1993 
  {\sl Phys. Rev.} C {\bf 48}, 2731
\bibitem{pan84} 
   Pandharipande V R, Papanicolas C N and Wmbach J 1984
   {\sl Phys. Rev. Lett.} {\bf 53}, 1133
\bibitem{ben90} 
   Benhar O, Fabrocini A and Fantoni S 1990 
   {\sl Phys. Rev.} C {\bf 41}, R24
\bibitem{fro83}
   Frois B {\sl et al} 1983 
   {\sl Nucl. Phys.} A {\bf A396} 409c
\bibitem{zam75}
   Zamick L, Klemt V  and  Speth J 1985
   {\sl Nucl. Phys.} A {\bf 245} 365 
\bibitem{kle76}
   Klemt V and Speth J 1976
   {\sl Zeit. Phys.} A {\bf 278} 59
\bibitem{kuo71}
   Kuo T T S  and Herling G H  1971
   {\sl Naval Research Laboratory Memorandum Report} {\bf 2258} 
\bibitem{dej87}
De Jager C W and De Vries C 1987
{\sl  At. Data Nucl. Data Tables}
{\bf 36} 495
\bibitem{hei69}
 Heisenberg J {\sl et al} 1969
 {\sl Phys. Rev. Lett.} {\bf 23} 1402 \\
 Eutener M {\sl et al} 1976
 {\sl Phys. Rev. Lett.} {\bf 36} 129 \\
 Frois B {\sl et al} 1977
 {\sl Phys. Rev. Lett.} {\bf 38} 152
\bibitem{fan84}
Fantoni S and Pandharipande V R
{\sl Nucl. Phys.} A {\bf 427} 473
\bibitem{ben89}
Benhar O, Fabrocini A and Fantoni S 1989
{\sl Nucl. Phys.} A {\bf 505} 267
%
\end{thebibliography}
\end{document}